\documentclass[prd,twocolumn,floatfix,amsmath,nofootinbib,amssymb,floatfix]{revtex4}
\usepackage{graphicx,color,dcolumn,booktabs,bm,multirow}
\usepackage{longtable,lscape}
\usepackage{txfonts}
\usepackage{overpic}
\usepackage{amssymb}
\usepackage{indentfirst}
\usepackage{feynmf}   
\usepackage{slashed}  
\usepackage{cases}
\usepackage{color}
\usepackage{multirow}
\usepackage{epstopdf}
\usepackage[colorlinks, citecolor=blue, linkcolor=blue, urlcolor=blue]{hyperref}

\begin{document}
\title{Strong decays of the possible $D^{*}K$ and $\bar{D}^{*}K$ molecules}
\author{Zi-Li Yue$^{1,2}$}
\email{yuezili@seu.edu.cn}
\author{Dian-Yong Chen$^{1,3}$\footnote{Corresponding author}} \email{chendy@seu.edu.cn}
\author{Elena Santopinto$^2$}
\email{elena.santopinto@ge.infn.it}
\affiliation{
 $^{1}$ School of Physics, Southeast University,  Nanjing 210094, China}
\affiliation{
 $^{2}$ INFN, Sezione di Genova, Via Dodecaneso 33, 16146 Genova, Italy}
\affiliation{$^3$ Lanzhou Center for Theoretical Physics, Lanzhou University, Lanzhou 730000, P. R. China}
\begin{abstract}
Inspired by the rich spectrum of structures near the $D^{(*)}K^{(*)}/\bar{D}^{(*)}K^{(*)}$ thresholds, we postulate the existence of $S$-wave $D^{*}K$ and $\bar{D}^{*}K$ molecules, denoted as $T_{c\bar{s}1}^{f}(2460)$ with $I=0$ (corresponding to $D_{s1}(2460)$), $T_{c\bar{s}1}^{a}(2470)$ with $I=1$, $T_{\bar{c}\bar{s}1}^{f}(2460)$ with $I=0$, and $T_{\bar{c}\bar{s}1}^{a}(2470)$ with $I=1$, respectively. Using an effective Lagrangian approach, we investigate the strong decays of these molecular states. Our estimates indicate that the width of $T_{c\bar{s}1}^{a}(2470)$ is roughly three orders of magnitude larger than that of $T_{c\bar{s}1}^{f}(2460)$, while the widths of $T_{\bar{c}\bar{s}1}^{a/f}(2470)$ are of the same orders as that of $T_{c\bar{s}1}^{f}(2460)$. The present study may provide valuable clues for the experimental search for these molecular candidates. 

\end{abstract}

\maketitle

\section{Introduction}\label{sec:1}

The past two decades have witnessed the discovery of new hadronic states beyond the conventional $q\bar{q}$ mesons and $qqq$ baryons (examples of experimental observations can be found in Refs.~\cite{Belle:2003nnu,Belle:2009and,Belle:2004lle,BESIII:2013ris,Belle:2013yex,BaBar:2003oey,LHCb:2024iuo,CLEO:2003ggt,LHCb:2020bls,LHCb:2024xyx,LHCb:2020pxc,LHCb:2022sfr,LHCb:2022lzp,D0:2016mwd,D0:2017qqm,LHCb:2015yax,LHCb:2016ztz}). In the sector of mesons with a single heavy flavor, the $D_{s0}^{*}(2317)$ and $D_{s1}(2460)$ are among the most prominent states. In 2003, the $D_{s0}^{*}(2317)$ was first discovered by the BaBar Collaboration in the $D_{s}^{+}\pi^{0}$ invariant mass spectrum from $e^{+}e^{-}$ annihilation data at energies near 10.6 GeV~\cite{BaBar:2003oey}. Subsequently, the CLEO Collaboration confirmed the discovery of $D_{s0}^{*}(2317)$ and further reported the observation of $D_{s1}(2460)$ in the $D_{s}^{*+}\pi^{0}$ invariant mass distribution with a statistical significance of $5.9\sigma$~\cite{CLEO:2003ggt}. These two states have been further observed in different processes by Belle and BaBar Collaborations~\cite{Belle:2003kup,Belle:2003guh,BaBar:2003cdx,BaBar:2004yux,BaBar:2006eep}.  The PDG average masses of $D_{s0}^{*}(2317)$ and $D_{s1}(2460)$ are $2317.8\pm0.5~\mathrm{MeV}$ and $2459.5\pm0.6~\mathrm{MeV}$, respectively~\cite{ParticleDataGroup:2024cfk}.

The $D_{s0}^{*}(2317)$ and $D_{s1}(2460)$ can be naturally interpreted as $c\bar{s}$ mesons with orbital angular momentum $L=1$ and spin-parity $J^{P}=0^{+}$ and $J^{P}=1^{+}$, respectively~\cite{CLEO:2003ggt,ParticleDataGroup:2024cfk}. However, the observed masses of the $D_{s0}^{*}(2317)$ and $D_{s1}(2460)$ are approximately $160~\mathrm{MeV}$ and $100~\mathrm{MeV}$ below their quark model predictions~\cite{Godfrey:1985xj}, respectively, which makes the interpretation of $D_{s0}^{*}(2317)$ and $D_{s1}(2460)$ as conventional $P$-wave charm-strange mesons questionable~\cite{Godfrey:2003kg,Song:2015nia,Colangelo:2004vu}. Given that their masses are close to the $DK$ and $D^{*}K$ thresholds, the corresponding molecular interpretations have been proposed. Consequently, numerous theoretical studies on their mass spectra~\cite{Cleven:2010aw,Guo:2006rp,Feng:2012zze,Xie:2010zza,Zhang:2006ix,Guo:2006fu,Barnes:2003dj,Chen:2004dy}, decay properties~\cite{Cleven:2014oka,Xiao:2016hoa,Faessler:2007gv,Faessler:2007us,Fu:2021wde}, and production mechanisms~\cite{Sakai:2017hpg,Liu:2022dmm,Liu:2023cwk,Guo:2014ppa,Albaladejo:2016hae,Sun:2015uva,Zhu:2019vnr,Navarra:2015iea,Albaladejo:2015kea} have been conducted within the $DK/D^{*}K$ molecular framework. In addition, the $DK$ and $D^{*}K$ coupled-channel effects have been introduced to understand the low-mass puzzle in Refs.~\cite{Ortega:2016mms,Ortega:2017zqu,Albaladejo:2018mhb,Liu:2009uz,Zhang:2024usz,Badalian:2007yr,Guo:2007up,Hwang:2004cd,Simonov:2004ar,Zhou:2011sp}.

Recently, the LHCb Collaboration observed two new full-open-flavor states, $T_{c\bar{s}}^{++}~(c\bar{d}u\bar{s})$ and $T_{c\bar{s}}^{0}~(c\bar{u}d\bar{s})$, in the $D_{s}^{+}\pi^{\pm}$ invariant mass spectrum from of $D_{s1}(2460)^{+}\to D_{s}^{+}\pi^{+}\pi^{-}$~\cite{LHCb:2024iuo}. The mass of $T_{c\bar{s}}^{++/0}$ is measured to be $m=2327\pm13\pm13~\mathrm{MeV}$, which lies near the $DK$ threshold, and the preferred assignment for the quantum numbers of $T_{c\bar{s}}^{++/0}$ is $I(J^{P})=1(0^{+})$~\cite{LHCb:2024iuo}. Both lattice QCD calculations~\cite{Gregory:2025ium} and an amplitude analysis of $D_{s1}(2460)^{+}\to D_{s}^{+}\pi^{+}\pi^{-}$ based on final-state interactions~\cite{Wang:2024fsz,Roca:2025lij} support the molecular interpretation of $T_{c\bar{s}0}(2327)$. Therefore, this state can be considered as the isospin vector partner of $D_{s0}^{*}(2317)$. In addition, the double-bump structure in the dipion invariant mass distributions of the $D_{s1}(2460)^{+}\to D_{s}^{+}\pi^{+}\pi^{-}$, predicted within the $D_{s1}(2460)$ molecule picture~\cite{Tang:2023yls}, has been experimentally confirmed~\cite{LHCb:2024iuo}, which further support the $D^\ast K$ molecular interpretation of $D_{s1}(2460)$.

\begin{figure}[t]
\centering
\includegraphics[width=8cm]{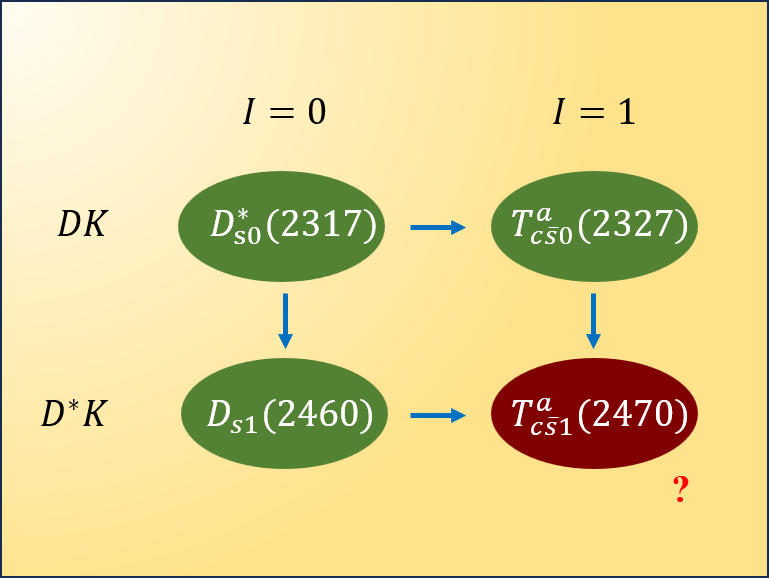}
\caption{Comparison of the $DK$ and $D^{*}K$ molecular states with different isospin. Similar to the case of $I=0$, the mass gap between $T_{c\bar{s}1}^a$ and $T_{c\bar{s}0}^a (2327)$ is approximately that of $D^\ast$ and $D$, indicating that the mass of $T_{c\bar{s}1}^a$ should be around 2470 MeV.} 
\label{Fig:Comp}
\end{figure}

As shown in Fig.~\ref{Fig:Comp}, the $DK$ molecular candidates $D_{s0}^{*}(2317)$ and its isospin vector partner $T_{c\bar{s}0}^{a}(2327)$ have been observed near the $DK$ threshold. The $D^{*}K$ molecular candidate $D_{s1}(2460)$ has also been observed near the $D^{*}K$ threshold. The mass difference between $D_{s1}(2460)$ and $D_{s0}^{*}(2317)$ approximately satisfies $m_{D_{s1}(2460)}-m_{D_{s0}^{*}(2317)} \approx m_{D^{*}}-m_{D}$, which can be naturally explained in the molecular framework. These observations indicate the existence of an isospin vector partner of $D_{s1}(2460)$. Following the LHCb naming convention, the $D^\ast K$ molecular states with $I=0$ and $I=1$ are named $T_{c\bar{s}1}^f$ and $T_{c\bar{s}1}^a$, respectively, while the state $T_{c\bar{s}1}^f$ corresponds to $D_{s1}(2460)$.

In addition, near the threshold of $D^{*}K^{*}$, two states, $T_{c\bar{s}0}^{a}(2900)^{++}~(c\bar{s}u\bar{d})$ and $T_{c\bar{s}0}^{a}(2900)^{0}(c\bar{s}\bar{u}d)$ have been discovered in the $D_{s}^{+}\pi^{\pm}$ invariant mass spectrum in the $B$ meson decay process by the LHCb Collaboration~\cite{LHCb:2022sfr,LHCb:2022lzp}. The resonance parameters of these two states are consistent with each other within experimental uncertainties, indicating them to be two of the isospin triplet states. In the same energy range, the LHCb Collaboration has also reported a state $T_{\bar{c}\bar{s}0}^{*}(2870)^{0}~(\bar{c}du\bar{s})$ (PDG name) in the $D^{-}K^{+}$ invariant mass spectrum of the $B^+ \to D^+ D^- K^+$ process~\cite{LHCb:2024xyx,LHCb:2020bls,LHCb:2020pxc}~\footnote{It's worth mentioning that another states $T_{\bar{c}\bar{s}1}^{*}(2900)^{0}~(\bar{c}du\bar{s})$ with spin one was also reported in the $D^{-}K^{+}$ invariant mass spectrum of the $B^+ \to D^+ D^- K^+$ process~\cite{LHCb:2020bls,LHCb:2020pxc}. In 2025, the LHCb Collaboration analyzed the amplitude of $B^{-}\to D^{-}D^{0}K_{S}^{0}$ and confirmed the observation of $T_{\bar{c}\bar{s}0}^{*}(2870)^{0}$ with a significance of $5.3~\sigma$, whereas no significant signal for $T_{\bar{c}\bar{s}1}^{*}(2900)^{0}$ was observed~\cite{LHCb:2024xyx}.}. Since $T_{c\bar{s}0}^a(2900)$ and $T_{\bar{c}\bar{s}0} (2870)$ are just close to the thresholds of $D^\ast K^\ast$ and $\bar{D}^\ast K^\ast$, respectively, the corresponding molecular explanations have been proposed~\cite{Liu:2020nil,Molina:2020hde,He:2020btl,Hu:2020mxp,Agaev:2020nrc,Mutuk:2020igv,Huang:2020ptc,Yue:2022mnf,Chen:2022svh,Agaev:2022eyk,Ke:2022ocs,Duan:2023lcj,Wang:2023hpp,Yue:2023qgx,Huang:2023fvj,Yu:2023avh,Yang:2024coj}. Similar to the case of $T_{c\bar{s}0}^a(2900)$ and $T_{\bar{c}\bar{s}0} (2870)$, the existence of $D^\ast K$ molecular state may also indicate possible molecular states near $\bar{D}^\ast K$ threshold, which can be named $T_{\bar{c}\bar{s}1}^f$ and $T_{\bar{c}\bar{s}1}^a$ for isospin singlet and triplets, respectively. In our previous work~\cite{Yue:2025wcl}, we investigated the strong decays of $D_{s0}^{*}(2317)$ and $T_{c\bar{s}0}^{a}(2327)$ within the $DK$ molecular framework using an effective Lagrangian approach. The estimated results are in good agreement with experimental measurements. Following this line of research, we study the strong decay properties of $T_{c\bar{s}1}^{a/f}/T_{\bar{c}\bar{s}1}^{a/f}$ in the $D^{*}K/\bar{D}^{*}K$ molecular framework, which may offer valuable clues for the experimental search of these states.

The rest of this work is organized as follows. The hadronic molecular structures of $T_{c\bar{s}1}^{f/a}$ and $T_{\bar{c}\bar{s}1}^{f/a}$ are introduced in Sec.~\ref{sec:2}. The strong decays of these molecular states are evaluated in Sec.~\ref{sec:3}. Numerical results and corresponding discussions are presented in Sec.~\ref{sec:4}, and a brief summary is given in the last section.

\section{Hadronic molecular structure}
\label{sec:2}

In the present work, $T_{c\bar{s}1}^{f/a}$ and $T_{\bar{c}\bar{s}1}^{f/a}$ are considered as $S$-wave $D^{*}K$ and $\bar{D}^{*}K$ molecules with isospin $I=0$ and $I=1$, respectively. The molecular structures for $T_{c\bar{s}1}^{f/a+}$ and $T_{\bar{c}\bar{s}1}^{f/a0}$ are

\begin{eqnarray}
\left|T_{c\bar{s}1}^{f+}\right>&=&\left|D^{*+}K^{0}\right>+\left|D^{*0}K^{+}\right>,\nonumber\\
\left|T_{c\bar{s}1}^{a+}\right>&=&\left|D^{*+}K^{0}\right>-\left|D^{*0}K^{+}\right>,\nonumber
\end{eqnarray}
\begin{eqnarray}
\left|T_{\bar{c}\bar{s}1}^{f0}\right>&=&\left|D^{*-}K^{+}\right>-\left|\bar{D}^{*0}K^{0}\right>,\nonumber\\
\left|T_{\bar{c}\bar{s}1}^{a0}\right>&=&\left|D^{*-}K^{+}\right>+\left|\bar{D}^{*0}K^{0}\right>,
\end{eqnarray}
respectively,

The effective Lagrangians describing the interactions between $T_{c\bar{s}1}^{f/a+}/T_{\bar{c}\bar{s}1}^{f/a0}$ and their components are,
\begin{eqnarray}
\label{eq:mo}
\mathcal{L}_{T_{c\bar{s}1}^{f/a+}}(x)&=&g_{T_{c\bar{s}1}^{f/a+}}T_{c\bar{s}0}^{f/a+\mu}(x) \int dy\Phi(y^2)\nonumber\\
&\times&\Big(D^{*+}_{\mu}(x+\omega_{K}y)K^{0}(x-\omega_{D^{*}}y)\nonumber\\
&\pm&D^{*0}_{\mu}(x+\omega_{K}y)K^{+}(x-\omega_{D^{*}}y)\Big)+\mathrm{h.c.},\nonumber\\
\label{eq:mo2}
\mathcal{L}_{T_{\bar{c}\bar{s}1}^{f/a0}}(x)&=&g_{T_{\bar{c}\bar{s}1}^{f/a0}}T_{\bar{c}\bar{s}1}^{f/a0\mu}(x)\int dy\Phi(y^2)\nonumber\\
&\times&\Big(D^{*-}_{\mu}(x+\omega_{K}y)K^{+}(x-\omega_{\bar{D}^{*}}y)\nonumber\\
&\mp&\bar{D}^{*0}_{\mu}(x+\omega_{K}y)K^{0}(x-\omega_{\bar{D}^{*}}y)\Big)+\mathrm{h.c.},
\end{eqnarray}
where $\omega_{i}=m_{i}/(m_{i}+m_{j})$ are the kinematical parameters, and $x$ and $y$ are the center-of-mass coordinate and the relative coordinate, respectively. The correlation function $\Phi(y^{2})$ corresponds to the wave function of $D^{*}K/\bar{D}^{*}K$. The Fourier transformation of $\Phi(y^{2})$ is
\begin{equation}
\Phi(y^{2})=\int \frac{d^{4}q}{(2\pi)^{4}}e^{-ipy}\tilde\Phi\left(-p^{2}\right).
\label{eq:fourier}
\end{equation}
The choice for $\tilde\Phi(-p^{2})$ should not only describe the inner structure of the $D^{*}K/\bar{D}^{*}K$ molecular states, but also fall off fast enough in the ultraviolet region of the Euclidean space. Here, we employ the correlation function in the Gaussian form~\cite{Faessler:2007gv,Faessler:2007us,Faessler:2008vc,Chen:2016byt,Dong:2017gaw,Gutsche:2010jf}, which is
\begin{equation}
\tilde\Phi(p_E^2)=\mathrm{exp}\left(-p_E^2/\Lambda^2\right),
\end{equation}
where $p_E=q_{E}-\omega_{DK}P_{E}$ represents the Euclidean Jacobi momentum and $\Lambda$ is a model parameter that parametrizes the distribution of the components inside the molecules.

The coupling constants $g_{T_{c\bar{s}1}^{f/a+}}$ and 
$g_{T_{\bar{c}\bar{s}1}^{f/a0}}$ in Eq.~(\ref{eq:mo}) can be determined by Weinberg's compositeness condition~\cite{Weinberg:1962hj, Salam:1962ap,vanKolck:2022lqz},
\begin{eqnarray}
\label{eq:cc}
Z&=&1-\Pi^{\prime}(m^{2})=0, 
\end{eqnarray}
where $\Pi^{\prime}$ is the derivative of the transverse part of the mass operator $\Pi^{\mu\nu}$,
\begin{eqnarray}
\Pi^{\mu\nu}(p)=g_{\perp}^{\mu\nu}\Pi(p^{2})+\frac{p^{\mu}p^{\nu}}{p^{2}}\Pi^{L}(p^{2}),
\end{eqnarray}
with $g_{\perp}^{\mu\nu}=g^{\mu\nu}-p^{\mu}p^{\nu}/p^{2}$ and $g_{\perp}^{\mu\nu}p_{\mu}=0$. $\Pi(p^2)$ and $Pi^L(p^2)$ are the conventional transverse and longitudinal component of the mass operator, respectively.

\begin{figure}[t]
\begin{tabular}{cc}
 \centering
  \includegraphics[width=4.0cm]{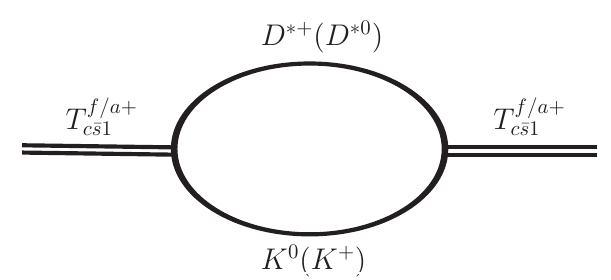}&
 \includegraphics[width=4.0cm]{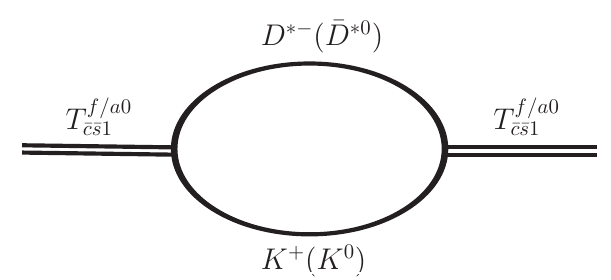}\\
 \\
$(a)$&$(b)$\\
 \end{tabular}
\caption{The mass operators of $T_{c\bar{s}1}^{f/a+}$ (diagram (a)) and $T_{\bar{c}\bar{s}1}^{f/a0}$ (diagram (b)) in the molecular framework.\label{fig:mo}}
\end{figure}

Based on the effective Lagrangians presented in Eq.~(\ref{eq:mo}), the concrete forms of the mass operators of $T_{c\bar{s}1}^{f/a+}$ and $T_{\bar{c}\bar{s}1}^{f/a0}$ corresponding to Fig.~\ref{fig:mo} can be written as
\begin{eqnarray}
  \label{eq:Pi}
\Pi^{\mu\nu}(m_{T_{c\bar{s}1}^{f/a+}}^{2})&=&2g_{T_{c\bar{s}1}^{f/a+}}^{2}\int\frac{d^{4}q}{(2\pi)^{4}}\tilde{\Phi}^{2}\left[-(q-\omega_{D^{*}}P)^{2},\Lambda^{2}\right]\nonumber\\
&\times&\frac{1}{(p-q)^{2}-m_{K}^2}\frac{-g^{\mu\nu}+q^{\mu}q^{\nu}/m_{D^{*}}^{2}}{q^{2}-m_{D^{*}}^{2}},\nonumber\\
\Pi^{\mu\nu}(m_{T_{\bar{c}\bar{s}1}^{f/a0}}^{2})&=&2g_{T_{\bar{c}\bar{s}1}^{f/a0}}^{2}\int\frac{d^{4}q}{(2\pi)^{4}}\tilde{\Phi}^{2}\left[-(q-\omega_{\bar{D}^{*}}P)^{2},\Lambda^{2}\right]\nonumber\\
&\times&\frac{1}{(p-q)^{2}-m_{K}^2}\frac{-g^{\mu\nu}+q^{\mu}q^{\nu}/m_{\bar{D}^{*}}^{2}}{q^{2}-m_{\bar{D}^{*}}^{2}},
\end{eqnarray}
where the factor $2$ comes from the isospins of the molecular components.

\section{Strong decays of $T_{c\bar{s}1}^{f/a+}$ and $T_{\bar{c}\bar{s}1}^{f/a0}$}
\label{sec:3}

\begin{figure*}[htb]
\begin{tabular}{ccc}
  \centering
\includegraphics[width=4.2cm]{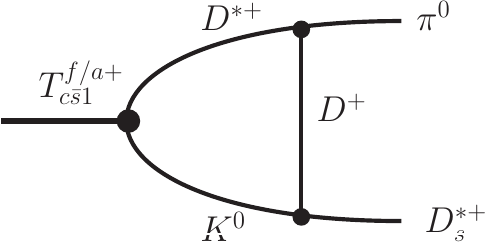}&
\includegraphics[width=4.2cm]{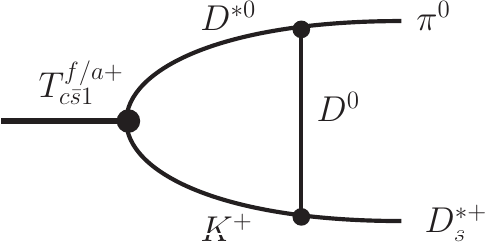}&
\includegraphics[width=4.2cm]{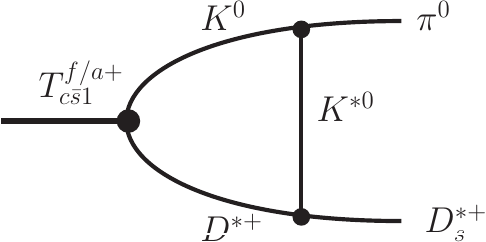}\\
 \\
 $(a)$ & $(b)$ & $(c)$\\ \\
\includegraphics[width=4.2cm]{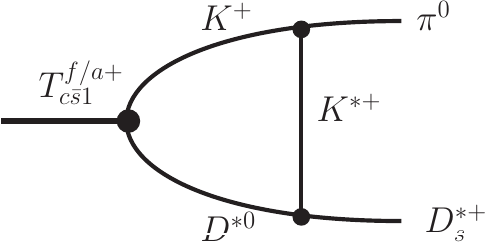}&
\includegraphics[width=4.2cm]{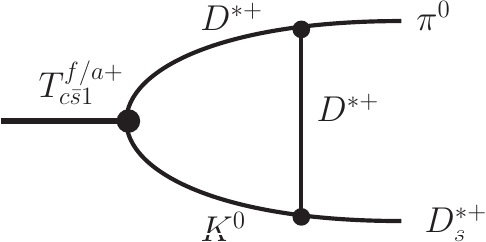}&
\includegraphics[width=4.2cm]{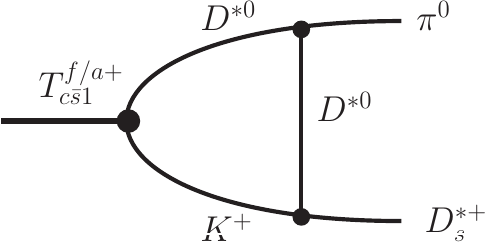}\\
 \\
$(d)$ & $(e)$ & $(f)$  \\ \\
\end{tabular}
\caption{Typical diagrams contributing to $T_{c\bar{s}1}^{f/a+}\to D_{s}^{*+}\pi^{0}$ at the hadron level. The diagrams contributing to the isospin-violating process $T_{c\bar{s1}}^{f+} \to D_{s0}^{\ast +} \pi^0$ induced by $\eta-\pi^0$ mixing are not displayed here.}
\label{fig:Decay}
\end{figure*}

\begin{figure*}[t]
\begin{tabular}{ccc}
  \centering
\includegraphics[width=4.2cm]{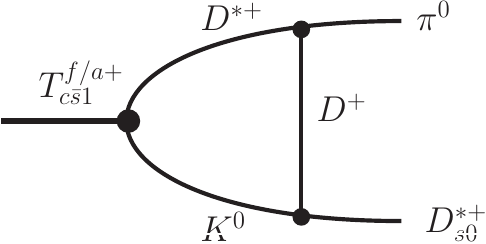}&
\includegraphics[width=4.2cm]{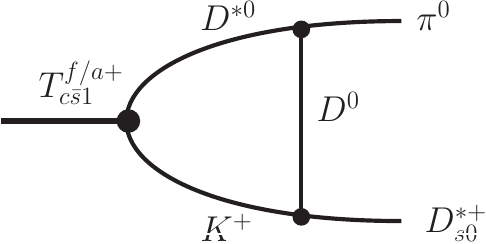}&
\includegraphics[width=4.2cm]{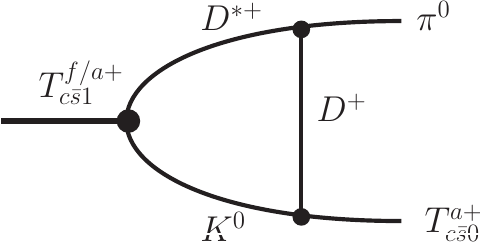}\\
 \\
 $(1)$ & $(2)$ & $(3)$\\ 
 \\
\includegraphics[width=4.2cm]{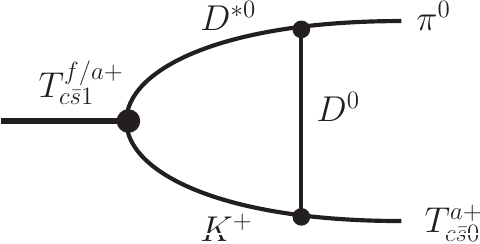}&
\includegraphics[width=4.2cm]{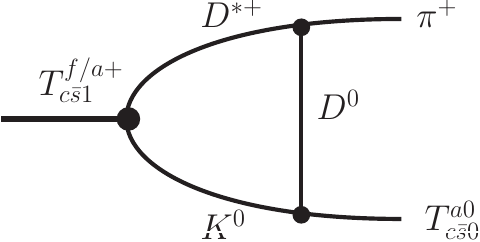}&
\includegraphics[width=4.2cm]{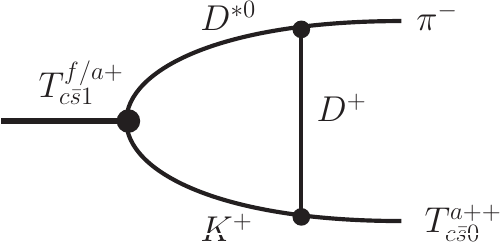}\\
 \\
 $(4)$&$(5)$ & $(6)$
\end{tabular}
\caption{Typical diagrams contributing to $T_{c\bar{s}1}^{f/a+}\to D_{s0}^{*+}\pi^{0}$ (diagrams (1) and (2)) and $T_{c\bar{s}1}^{f/a+}\to T_{c\bar{s}0}^{a}\pi$ (diagrams (3)-(6)) at the hadron level. The diagrams contributing to the isospin-violating process $T_{c\bar{s1}}^{f+} \to D_{s0}^{\ast +} \pi^0$ induced by $\eta-\pi^0$ mixing are not displayed here.}
\label{fig:moDecay}
\end{figure*}

In this section, we investigate the strong decay behaviors of $T_{c\bar{s}1}^{f/a+}$ and $T_{\bar{c}\bar{s}1}^{f/a0}$ in the $D^{*}K$ and $\bar{D}^\ast K$ molecular framework. For $T_{c\bar{s}1}^{f/a+}$ states, the possible strong decay channels include $D_{s}^{*+}\pi^{0},~D_{s0}^{*+}\pi^{0},~T_{c\bar{s}0}^{a}\pi$ (including $T_{c\bar{s}0}^{a0}\pi^{+}$, $T_{c\bar{s}0}^{a+}\pi^{0}$, and $T_{c\bar{s}0}^{a++}\pi^{-}$). The corresponding decay diagrams are collected in Fig.~\ref{fig:Decay} and Fig.~\ref{fig:moDecay}. Regarding $T_{\bar{c}\bar{s}1}^{a0}$ and $T_{\bar{c}\bar{s}1}^{f0}$, their decays into charm-strange and light mesons or charm and strange mesons are forbidden due to their quark components. The possible strong decay processes may include $T_{\bar{c}\bar{s}1}^{a0}\to T_{\bar{c}\bar{s}0}^{f0}\pi^{0}$, $T_{\bar{c}\bar{s}1}^{a0}\to T_{\bar{c}\bar{s}0}^{a0}\pi^{0}$, $T_{\bar{c}\bar{s}1}^{f0}\to T_{\bar{c}\bar{s}0}^{f0}\pi^{0}$, and $T_{\bar{c}\bar{s}1}^{f0}\to T_{\bar{c}\bar{s}0}^{a0}\pi^{0}$ if there exist $DK/\bar{D}K$ molecular states. In the following, we mainly focus on the decay processes of $T_{c\bar{s}1}^{f/a+}$ states.

\subsection{Effective Lagrangian}
To evaluate the diagrams in Fig.~\ref{fig:Decay} and Fig.~\ref{fig:moDecay}, we employ the SU(4) Lagrangians to describe the interactions between pseudoscalar and vector mesons, with the vector mesons introduced within the massive Yang-Mills framework~\cite{Faessler:2007gv,Haglin:2000ar,Haglin:1999xs,Lin:1999ad,Azevedo:2003qh,Kaymakcalan:1984bz,Gomm:1984at,Yue:2025wcl}. The concrete effective Lagrangians read,
\begin{eqnarray}
\mathcal{L}_{D^{*}D\pi}&=&ig_{D^{*}D\pi}\left(\bar{D}\vec{\tau}\cdot\partial_{\mu}\vec{\pi}-\vec{\tau}\cdot\vec{\pi}\partial_{\mu}\bar{D}\right)D^{*\mu}+\mathrm{h.c.}\nonumber\\
\mathcal{L}_{D^{*}D\eta}&=&ig_{D^{*}D\eta}\Big(\eta\partial_{\mu}D-D\partial_{\mu}\eta\Big)\bar{D}^{*\mu}+\mathrm{h.c.}\nonumber\\
\mathcal{L}_{K^{*}K\pi}&=&ig_{K^{*}K\pi}\left(K\vec{\tau}\cdot\partial_{\mu}\vec{\pi}-\vec{\tau}\cdot\vec{\pi}\partial_{\mu}K\right)\bar{K}^{*\mu}+\mathrm{h.c.}\nonumber\\
\mathcal{L}_{K^{*}K\eta}&=&ig_{K^{*}K\eta}\Big(K\partial_{\mu}\eta-\eta\partial_{\mu}K\Big)\bar{K}^{*\mu}+\mathrm{h.c.}\nonumber\\
\mathcal{L}_{D_{s}^{*}DK}&=&ig_{D_{s}^{*}DK}\left(K\partial_{\mu}D-D\partial_{\mu}K\right)\bar{D}_{s}^{*}+\mathrm{h.c.}\nonumber\\
\mathcal{L}_{D_{s}^{*}D^{*}K^{*}}&=&ig_{D_{s}^{*}D^{*}K^{*}}\left(\partial_{\mu}K^{*}_{\nu}\left(\bar{D}_{s}^{*\mu}D^{*\nu}-\bar{D}_{s}^{*\nu}D^{*\mu}\right)\right.\nonumber\\
&+&\left.\partial_{\mu}\bar{D}_{s\nu}^{*}\left(D^{*\mu}K^{*\nu}-D^{*\nu}K^{*\mu}\right)\right.\nonumber\\
&+&\left.\partial_{\mu}D^{*}_{\nu}\left(\bar{D}_{s}^{*\nu}K^{*\mu}-\bar{D}_{s}^{*\mu}K^{*\nu}\right)\right)+\mathrm{h.c.}\nonumber\\
\mathcal{L}_{D^{*}D^{*}\pi}&=&g_{D^{*}D^{*}\pi}\epsilon_{\mu\nu\alpha\beta}\left(\partial^{\mu}\bar{D}^{*\nu}\partial^{\alpha}D^{*\beta}\vec{\tau}\cdot\vec{\pi}\right)+\mathrm{h.c.}\nonumber\\
\mathcal{L}_{D^{*}D^{*}\eta}&=&g_{D^{*}D^{*}\eta}\epsilon_{\mu\nu\alpha\beta}\left(\partial^{\mu}\bar{D}^{*\nu}\partial^{\alpha}D^{*\beta}\eta\right)+\mathrm{h.c.}\nonumber\\
\mathcal{L}_{D_{s}^{*}D^{*}K}&=&g_{D_{s}^{*}D^{*}K}\epsilon_{\mu\nu\alpha\beta}\left(\partial^{\mu}\bar{D}_{s}^{*\nu}\partial^{\alpha}D^{*\beta}\right)+\mathrm{h.c.}
\label{eq:Lag-SU4}
\end{eqnarray}

Regarding the isospin-violating process $T_{c\bar{s}1}^{f+}\to D_{s}^{*+}\pi^{0}$, the contribution from $\eta$-$\pi^{0}$ mixing should also be considered. The effective Lagrangian related to $\eta$-$\pi^{0}$ mixing is~\cite{Gasser:1983yg,Gasser:1984gg,Pich:1995bw},
\begin{eqnarray}
\mathcal{L}_{\eta\pi^{0}}=\mu\frac{m_{d}-m_{u}}{\sqrt{3}}\pi^{0}\eta,
\end{eqnarray}
where $m_u$ and $m_d$ are the $u$ and $d$ current quark masses, and $\mu$ is the condensate parameter. The $D^{(*)}=\left(D^{(*)0},~D^{(*)+}\right)$ and $K^{(*)}=\left(K^{(*)+},~K^{(*)0}\right)^{T}$ denote the charm and strange doublets, respectively. Their charge-conjugate counterparts are denoted by $\bar{D}^{(*)}$ and $\bar{K}^{(*)}$, respectively.

\subsection{Decay amplitude}

With the effective Lagrangians in Eqs.~(\ref{eq:mo}) and~(\ref{eq:Lag-SU4}), the amplitudes corresponding to the diagrams in Fig.~\ref{fig:Decay}-($a$), ($c$), and ($e$) can be written as,
\begin{eqnarray}
i\mathcal{M}_{a}&=&i^{3}\int\frac{d^{4}q}{(2\pi)^{4}}\Big[g_{T_{c\bar{s}1}^{f/a+}}\tilde{\Phi}(-p_{12}^{2},\Lambda^{2})\epsilon_{\mu}(P)\Big]\nonumber\\
&\times&\Big[ig_{D^{*}D\pi}\left(iq^{\nu}-ip_{3}^{\nu}\right)\Big]\Big[ig_{DD_{s}^{*}K}\left(-iq^{\rho}+ip_{2}^{\rho}\right)\epsilon_{\rho}(p_{4})\Big]\nonumber\\
&\times&\frac{-g_{\mu\nu}+p_{1\mu}p_{1\nu}/m_{1}^{2}}{p_{1}^{2}-m_{1}^{2}}\frac{1}{p_{2}^{2}-m_{2}^{2}}\frac{1}{q^{2}-m_{q}^{2}},\nonumber
\end{eqnarray}

\begin{eqnarray}
i\mathcal{M}_{c}&=&i^{3}\int\frac{d^{4}q}{(2\pi)^{4}}\Big[g_{T_{c\bar{s}1}^{f/a+}}\tilde{\Phi}(-p_{12}^{2},\Lambda^{2})\epsilon_{\mu}(P)\Big]\nonumber\\
&\times&\Big[ig_{K^{*}K\pi}\left(-ip_{1}^{\alpha}-ip_{3}^{\alpha}\right)\Big]\Big[ig_{D^{*}D_{s}^{*}K^{*}}\left(-i\left(q^{\rho}g^{\beta\nu}-g^{\beta\rho}q^{\nu}\right)\right.\nonumber\\
&+&\left.i\left(p_{4}^{\nu}g^{\rho\beta}-g^{\rho\nu}p_{4}^{\beta}\right)-i\left(g^{\nu\rho}p_{2}^{\beta}-p_{2}^{\rho}g^{\nu\beta}\right)\right)\epsilon_{\rho}(p_{4})\Big]\nonumber\\
&\times&\frac{1}{p_{1}^{2}-m_{1}^{2}}\frac{-g_{\mu\nu}+p_{2\mu}p_{2\nu}/m_{2}^{2}}{p_{2}^{2}-m_{2}^{2}}\frac{-g_{\alpha\beta}+q_{\alpha}q_{\beta}/m_{q}^{2}}{q^{2}-m_{q}^{2}},\nonumber\\
i\mathcal{M}_{e}&=&i^{3}\int\frac{d^{4}q}{(2\pi)^{4}}\Big[g_{T_{c\bar{s}1}^{f/a+}}\tilde{\Phi}(-p_{12}^{2},\Lambda^{2})\epsilon_{\phi}(P)\Big]\nonumber\\
&\times&\Big[ig_{D^{*}D^{*}\pi}\epsilon_{\mu\lambda\alpha\beta}\left(-iq^{\mu}(-ip_{1}^{\alpha})\right)\Big]\Big[ig_{D^{*}D_{s}^{*}K}\epsilon_{\tau\rho\theta\sigma}\nonumber\\&\times&\left(ip_{4}^{\tau}(-iq^{\theta})\right)\epsilon_{\rho}(p_{4})\Big]\frac{-g_{\phi\beta}+p_{1\phi}p_{1\beta}/m_{1}^{2}}{p_{1}^{2}-m_{1}^{2}}\frac{1}{p_{2}^{2}-m_{2}^{2}}\nonumber\\
&\times&\frac{-g_{\lambda\sigma}+q_{\lambda}q_{\sigma}/m_{q}^{2}}{q^{2}-m_{q}^{2}},\nonumber
\end{eqnarray}

The amplitudes corresponding to diagrams ($b$), ($d$), and ($f$) in Fig.~\ref{fig:Decay} can be obtained by employing the following replacements,
\begin{eqnarray}
\mathcal{M}_{b}&=&\mathcal{M}_{a}\Big|_{D^{*+}\to D^{*0}, K^{0}\to K^{+}, D^{+}\to D^{0}},\nonumber\\
\mathcal{M}_{d}&=&\mathcal{M}_{c}\Big|_{K^{0}\to K^{+}, D^{*+}\to D^{*0}, K^{*0}\to K^{*+}},\nonumber\\
\mathcal{M}_{f}&=&\mathcal{M}_{e}\Big|_{D^{*+}\to D^{*0}, K^{0}\to K^{+}, D^{*+}\to D^{*0}}.
\end{eqnarray}

For the $T_{c\bar{s}1}^{f+}\to D_{s}^{*+}\pi^{0}$ process, the $\eta$-$\pi^{0}$ mixing cannot be neglected, the corresponding diagrams are very similar to those in Fig.~\ref{fig:Decay}, which can be obtained by employing the following substitutions,
\begin{eqnarray}
\mathcal{M}_{a}^{mix}&=&\mathcal{M}_{a}\frac{m_{d}-m_{u}}{m_{s}-m}\frac{\sqrt{3}}{4}\Big|_{g_{D^{*}D\pi}\to g_{D^{*}D\eta}},\nonumber\\
\mathcal{M}_{b}^{mix}&=&\mathcal{M}_{b}\frac{m_{d}-m_{u}}{m_{s}-m}\frac{\sqrt{3}}{4}\Big|_{g_{D^{*}D\pi}\to g_{D^{*}D\eta}},\nonumber\\
\mathcal{M}_{c}^{mix}&=&\mathcal{M}_{c}\frac{m_{d}-m_{u}}{m_{s}-m}\frac{\sqrt{3}}{4}\Big|_{g_{K^{*}K\pi}\to g_{K^{*}K\eta}},\nonumber\\
\mathcal{M}_{d}^{mix}&=&\mathcal{M}_{d}\frac{m_{d}-m_{u}}{m_{s}-m}\frac{\sqrt{3}}{4}\Big|_{g_{K^{*}K\pi}\to g_{K^{*}K\eta}},\nonumber\\
\mathcal{M}_{e}^{mix}&=&\mathcal{M}_{e}\frac{m_{d}-m_{u}}{m_{s}-m}\frac{\sqrt{3}}{4}\Big|_{g_{D^{*}D^{*}\pi}\to g_{D^{*}D^{*}\eta}},\nonumber\\
\mathcal{M}_{f}^{mix}&=&\mathcal{M}_{f}\frac{m_{d}-m_{u}}{m_{s}-m}\frac{\sqrt{3}}{4}\Big|_{g_{D^{*}D^{*}\pi}\to g_{D^{*}D^{*}\eta}},\label{Eq:etapimix}
\end{eqnarray}
with $m=(m_{u}+m_{d})/2$.

The pionic transitions from $T_{c\bar{s}1}^{f/a}$ to $D_{s0}^{*+}$ and $T_{c\bar{s}0}^{a}$ are shown in Fig.~\ref{fig:moDecay}. The amplitude corresponding to Fig.~\ref{fig:moDecay}-(1) reads,
\begin{eqnarray}
i\mathcal{M}_{1}&=&i^{3}\int\frac{d^{4}q}{(2\pi)^{4}}\Big[g_{T_{c\bar{s}1}^{f/a+}}\tilde{\Phi}(-p_{12}^{2},\Lambda^{2})\epsilon_{\mu}(P)\Big]\nonumber\\
&\times&\left[g_{D_{s0}^{*}}\tilde{\Phi}_{D_{s0}^{*}}(-p_{20}^{2},\Lambda_{D_{s1}}^{2})\right]\left[ig_{D^{*}D\pi}(iq^{\nu}-ip_{3}^{\nu})\right]\nonumber\\
&\times&\frac{-g_{\mu\nu}+p_{1\mu}p_{1\nu}/m_{1}^{2}}{p_{1}^{2}-m_{1}^{2}}\frac{1}{p_{2}^{2}-m_{2}^{2}}\frac{1}{q^{2}-m_{q}^{2}}.
\end{eqnarray}
where $D_{s0}^{*+}$ and $T_{c\bar{s}0}^{a}$ are considered as $S$-wave $DK$ molecular state with isospin equal to $0$ and $1$, respectively. The effective Lagrangians describing the interactions between these molecular states and their components are introduced in Eq.~(2) in \cite{Yue:2025wcl}. In the following estimations, we take the same $\Lambda$ for both $D^\ast K/\bar{D}^\ast K$ and $DK/\bar{D}K$ molecular states. Similar to the case of $T_{c\bar{s}1}^{f/a^+}\to D_s^{\ast} \pi^0$, one can obtain the amplitude corresponding to Fig~\ref{fig:moDecay}-(1) by, 
\begin{eqnarray}
	i \mathcal{M}_2= \mathcal{M}_1\Big|_{D^{\ast +} \to D^{\ast0}, K^0\to K^+ D^+ \to D^0 },
\end{eqnarray}
and by replacing the coupling constant $g_{D_{s0}^{\ast+} DK}$ in the amplitudes $\mathcal{M}_1$ and $\mathcal{M}_2$ with $g_{T_{c\bar{s}0^{a+} DK}}$, one can obtain the amplitudes for $T_{c\bar{s}1}^{f/a+} \to T_{c\bar{s}0}^{a+} \pi^0$, which are $\mathcal{M}_3$ and $\mathcal{M}_4$ corresponding Fig~\ref{fig:moDecay}-(3) and (4), respectively. In a similar way, one can obtain the amplitudes, $\mathcal{M}_5$ and $\mathcal{M}_6$, corresponding to Fig~\ref{fig:moDecay}-(5) and (6). Considering $\eta-\pi^0$ mixing, there are additional diagrams contributing to isospin violating process $T_{c\bar{s}1}^{f+} \to D_{s0}^{\ast +} \pi^0$, which can be obtained from $\mathcal{M}_1$ and $\mathcal{M}_2$ by performing the same substitutions as in Eq.~\eqref{Eq:etapimix}, and the corresponding amplitudes are named $\mathcal{M}_1^{mix}$ and $\mathcal{M}_2^{mix}$, respectively.

Then, the amplitudes for the considered strong decay processes of $T_{c\bar{s}1}^{f+}$ and $T_{c\bar{s}1}^{a+}$ are,
\begin{eqnarray}
\label{eq:total}
\mathcal{M}_{T_{c\bar{s}1}^{f+}\to D_{s}^{*+}\pi^{0}}&=&\mathcal{M}_{a}+\mathcal{M}_{c}+\mathcal{M}_{e}-\mathcal{M}_{b}-\mathcal{M}_{d}-\mathcal{M}_{f}\nonumber\\
&+&\mathcal{M}_{a}^{mix}+\mathcal{M}_{c}^{mix}+\mathcal{M}_{e}^{mix}+\mathcal{M}_{b}^{mix}+\mathcal{M}_{d}^{mix}+\mathcal{M}_{f}^{mix}\nonumber\\
\mathcal{M}_{T_{c\bar{s}1}^{f+}\to D_{s0}^{*+}\pi^{0}}
&=&\mathcal{M}_{1}-\mathcal{M}_{2}+\mathcal{M}_{1}^{mix}+\mathcal{M}_{2}^{mix}\nonumber\\
\mathcal{M}_{T_{c\bar{s}1}^{f+}\to T_{c\bar{s}0}^{a}\pi^{0}}&=&\mathcal{M}_{3}+\mathcal{M}_{4}\nonumber\\
\mathcal{M}_{T_{c\bar{s}1}^{f+}\to T_{c\bar{s}0}^{a}\pi^{+}}&=&\mathcal{M}_{5}\nonumber\\
\mathcal{M}_{T_{c\bar{s}1}^{f+}\to T_{c\bar{s}0}^{a}\pi^{-}}&=&\mathcal{M}_{6}\nonumber\\
\mathcal{M}_{T_{c\bar{s}1}^{a+}\to D_{s}^{*}\pi^{0}}&=&\mathcal{M}_{a}+\mathcal{M}_{c}+\mathcal{M}_{e}-\mathcal{M}_{b}-\mathcal{M}_{d}-\mathcal{M}_{f}\nonumber\\
\mathcal{M}_{T_{c\bar{s}1}^{a+}\to D_{s0}^{*+}\pi^{0}}&=&\mathcal{M}_{1}+\mathcal{M}_{2}\nonumber\\
\mathcal{M}_{T_{c\bar{s}1}^{a+}\to T_{c\bar{s}0}^{a}\pi}&=&\mathcal{M}_{3}+\mathcal{M}_{4}\nonumber\\
\mathcal{M}_{T_{c\bar{s}1}^{a+}\to T_{c\bar{s}0}^{a}\pi}&=&\mathcal{M}_{5}\nonumber\\
\mathcal{M}_{T_{c\bar{s}1}^{a+}\to T_{c\bar{s}0}^{a}\pi}&=&\mathcal{M}_{6}.
\end{eqnarray}

With the above amplitudes, the partial widths for the considered processes can be estimated by
\begin{eqnarray}
\Gamma_{A\to BC}=\frac{1}{3}\frac{1}{8\pi}\frac{|\vec{p}|}{m_{0}^{2}}\Big|\overline{\mathcal{M}_{A\to BC}}\Big|^{2},
\end{eqnarray}
where $m_{0}$ refers to the mass of the initial state, and $\vec{p}$ is the final-state momentum in the rest frame of the initial state. The factor $1/3$ comes from the average of the spin of the initial state.

\section{Numerical results and discussions}\label{sec3}
\label{sec:4}
\subsection{Coupling constants}

The couplings $g_{D^{*}D\pi}=6.22$ and $g_{K^{*}K\pi}=3.12$ are taken from the measured decay widths of the processes $D^{*+}\to D^{+}\pi^{0}$ and $K^{*0}\to K^{0}\pi^{0}$~\cite{ParticleDataGroup:2024cfk}, respectively. By using the QCD sum rule method, one has the coupling constants $g_{DD_{s}K^{*}}=g_{D^{*}D_{s}K}=2.02$ and $g_{D_{s}^{*}DK}=1.84$~\cite{Wang:2006ida,Bracco:2006xf}.  The relationship between $g_{D^{*}D^{*}\pi}$/$g_{K^{*}K\pi}$ and $g_{D^{*}D^{*}\eta}$/$g_{K^{*}K\eta}$ can be written in terms of leptonic decay constants as following,
\begin{eqnarray}
g_{D^{*}D\eta}=\frac{f_{\pi}}{f_{\eta}\sqrt{3}}g_{D^{*}D\pi},~~~~g_{K^{*}K\eta}=\frac{f_{\pi}\sqrt{3}}{f_{\eta}}g_{K^{*}K\pi}.
\label{eq:coup-2}
\end{eqnarray}

Regarding the couplings $g_{D^{*}D^{*}\pi}$, $g_{D^{*}D^{*}\eta}$, $g_{D_{s}^{*}D^{*}K^{*}}$ and $g_{D_{s}^{*}D^{*}K}$, they can be determined via the following relations~\cite{Lin:1999ad,Azevedo:2003qh}
\begin{eqnarray}
g_{D^{*}D\pi}&=&\frac{1}{4}g_{1},~~~~g_{D^{*}D^{*}\pi}=-\frac{1}{4}\frac{g_{1}^{2}N_{c}}{16\pi^{2}f_{\pi}},\nonumber\\
g_{D^{*}D\eta}&=&\frac{1}{4}g_{2},~~~~g_{D^{*}D^{*}\eta}=-\frac{1}{4}\frac{g_{2}^{2}N_{c}}{16\pi^{2}f_{\eta}},\nonumber\\
g_{D^{*}D_{s}K}&=&g_{D_{s}^{*}D^{*}K^{*}}=\frac{1}{4}g_{3},~~~~g_{D_{s}^{*}D^{*}K}=-\frac{1}{4}\frac{g_{3}^{2}N_{c}}{16\pi^{2}f_{K}}.
\label{eq:coup-1}
\end{eqnarray}
where the $f_{\pi}=132~\mathrm{MeV}$, $f_{\eta}=170~\mathrm{MeV}$ and $f_{K}=155.7~\mathrm{MeV}$ are the $\pi$, $\eta$ and $K$ decay constants, respectively~\cite{ParticleDataGroup:2024cfk}.

\begin{figure}[t]
  \centering
  \includegraphics[width=8.3 cm]{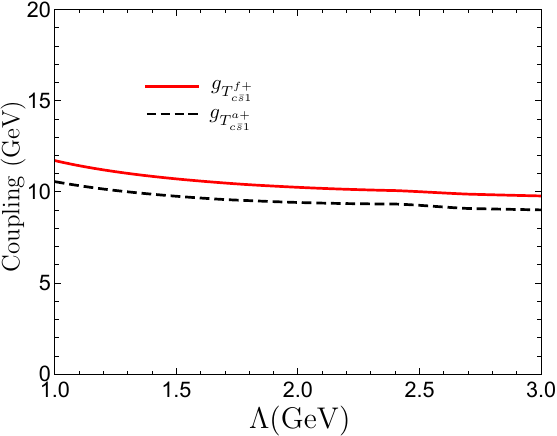}
  
\caption{(Color online.) The coupling constants $g_{T_{c\bar{s}1}^{f+}}$ (red line) and $g_{T_{c\bar{s}1}^{a+}}$ (black line) as functions of the model parameter $\Lambda$, where the mass of $T_{c\bar{s}1}^{a+}$ is taken to be $m(T_{c\bar{s}1}^{a+})=2.470~\mathrm{GeV}$.}\label{fig:couplingx}
\end{figure}

The coupling constants between the molecular states $T_{c\bar{s}1}^{f/a+}$, $D_{s0}^{*+}/T_{c\bar{s}0}^{a+}$, and their components $D^{*}K$ and $DK$ can be determined by using Weinberg's compositeness condition in Eq.~(\ref{eq:cc}). As shown in Eq.~\eqref{eq:Pi}, the coupling constants determined from Eq.~(\ref{eq:cc}) depend on the parameter $\Lambda$. Empirically, this model parameter should be of the order of $1~\mathrm{GeV}$. The couplings $g_{D_{s0}^{*+}}$ and $g_{T_{c\bar{s}0}^{a+}}$ have been discussed in our previous work~\cite{Yue:2025wcl}. Regarding $T_{c\bar{s}1}^{a+}$, we use the relation $m_{T_{c\bar{s}1}^{a}}-m_{T_{c\bar{s}0}^{a}(2327)} \approx m_{D^{*}}-m_{D}$ to estimate the mass of $T_{c\bar{s}1}^{a}$, with the central value of $m_{T_{c\bar{s}0}^{a}(2327)}$, the mass of $T_{c\bar{s}1}^{a}$ is estimated to be around 2.470 GeV. The couplings $g_{T_{c\bar{s}1}^{f+}}$ and $g_{T_{c\bar{s}1}^{a+}}$ as functions of the parameter $\Lambda$ are presented in Fig.~\ref{fig:couplingx}. With the parameter $\Lambda$ ranging from $1~\mathrm{GeV}$ to $3~\mathrm{GeV}$, the coupling constants $g_{T_{c\bar{s}1}^{f+}}$ and $g_{T_{c\bar{s}1}^{a+}}$ decrease from $11.7~\mathrm{GeV}$ to $9.78~\mathrm{GeV}$ and from $10.6~\mathrm{GeV}$ to $9.02~\mathrm{GeV}$, respectively. The values of these two coupling constants are very close to the coupling constants between $D_{s0}(2317)/T_{c\bar{s}}(2327)$ and their component $DK$~\cite{Yue:2025wcl}.

\subsection{Strong decays of $T_{c\bar{s}1}^{f+}$}
\begin{figure}[t]
  \centering
   
\includegraphics[width=8.3 cm]{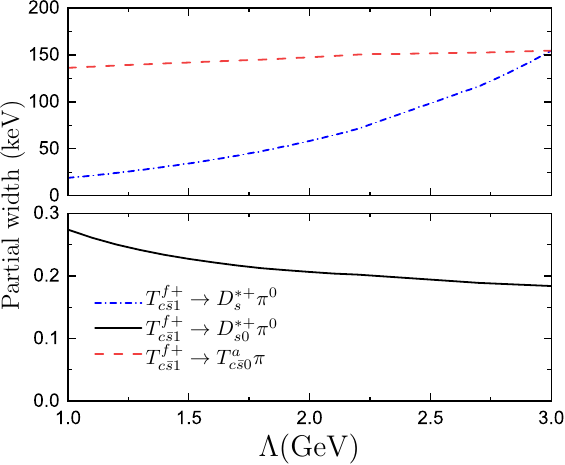}
\caption{(Color online) The partial decay widths of $T_{c\bar{s}1}^{f+}\to D_{s}^{*+}\pi^{0}$ and $D_{s0}^{*+}\pi^{0}$,  considering the $\eta$-$\pi^{0}$ mixing, and $T_{c\bar{s}1}^{f+}\to T_{c\bar{s}0}^{a}\pi$, as functions of the model parameter $\Lambda$s.}\label{fig:width1}
\end{figure}

In Fig.~\ref{fig:width1}, we present the partial decay widths of the processes $T_{c\bar{s}1}^{f+}\to D_{s}^{*+}\pi^{0}$, $D_{s0}^{*+}\pi^{0}$, and $T_{c\bar{s}0}^{a}\pi$ as functions of the model parameter $\Lambda$. The contribution from $\eta$-$\pi^{0}$ mixing has also been considered in the isospin-violating processes $T_{c\bar{s}1}^{f+}\to D_{s}^{*+}\pi^{0}$ and $T_{c\bar{s}1}^{f+}\to D_{s0}^{*+}\pi^{0}$. In the parameter range, the partial width of $T_{c\bar{s}1}^{f+}\to D_{s}^{*+}\pi^{0}$ increases from $18.98~\mathrm{keV}$ to $154.4~\mathrm{keV}$ with $\Lambda$ increasing from $1~\mathrm{GeV}$ to $3~\mathrm{GeV}$. As for $T_{c\bar{s}1}^{f+}\to D_{s0}^{*+}\pi^{0}$, this process is suppressed by both isospin violation and kinematic limitation since $m_{T_{cs1}^{f+}(2460)}-m_{D_{s0}^\ast (2317)} \approx m_{\pi^0}$. Our estimations indicate that the partial width of this process decreases from 0.27 keV to 0.18 keV with  $\Lambda$ increasing from $1~\mathrm{GeV}$ to $3~\mathrm{GeV}$

\begin{figure}[t]
  \centering
  \includegraphics[width=8.3 cm]{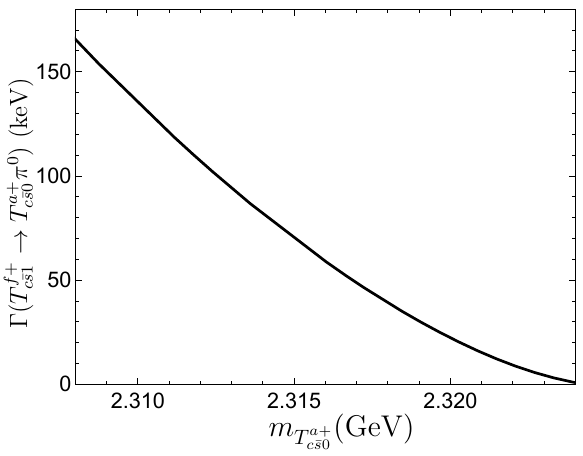}
\caption{The partial width of $T_{c\bar{s}1}^{f+}\to T_{c\bar{s}0}^{+}\pi^{0}$ as a function of the mass of $T_{c\bar{s}0}^{a+}$ when $\Lambda=1~\mathrm{GeV}$.}\label{fig:mass}
\end{figure}

Regarding the pionic transitions between molecular states $T_{c\bar{s}1}^{f+}(2460)$ and $T_{c\bar{s}0}^{a}(2327)$, this process is isospin conserved, but suppressed kinematically since the phase space of this process is very limited. Taking $m_{T_{c\bar{s}0}^{a+}}=2317~\mathrm{MeV}$, we find that the estimated partial width depends very weakly on the model parameter. In particular, the width of $T_{c\bar{s}1}^{f+}(2460) \to T_{c\bar{s}0}^a \pi$ (including $ T_{c\bar{s}0}^0 \pi^+$, $ T_{c\bar{s}0}^+ \pi^0$ and $ T_{c\bar{s}0}^{++} \pi^-$) is estimated to be $136.2 \sim 153.4~\mathrm{keV}$ in the considered parameter range. Moreover, in Fig.~\ref{fig:mass}, we present the partial width of $T_{c\bar{s}1}^{f+}\to T_{c\bar{s}0}^{a}\pi^0$ depending on the mass of $T_{c\bar{s}0}^{a}$ with $\Lambda=1.0 $ GeV, where the mass of $T_{c\bar{s}0}^{a}$ varies from the lower limit of $2.308~\mathrm{GeV}$ to $2.324~\mathrm{GeV}$. With the considered  mass range, the partial width of $T_{c\bar{s}1}^{f+}\to T_{c\bar{s}0}^{a}\pi^0$ decreases from 165 keV to 0.9 keV. 

In conclusion, the strong decay processes of $T_{c\bar{s}1}^{f+}$ that we have considered in the present article, are suppressed by isospin violation or kinematic limitation or both, leading to very small partial widths of these processes.

\subsection{Strong decays of $T_{c\bar{s}1}^{a+}$}
\begin{figure}[t]
  \centering
  \includegraphics[width=8.3 cm]{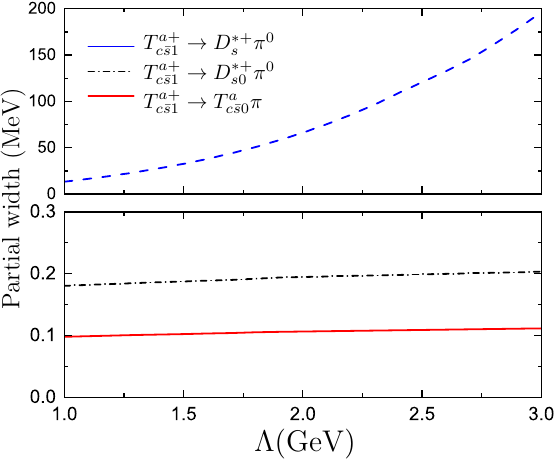}
\caption{(Color online) The partial decay widths of $T_{c\bar{s}1}^{a+}\to D_{s}^{*+}\pi^{0}$, $D_{s0}^{*+}\pi^{0}$, and $T_{c\bar{s}0}^{a}\pi$ as functions of the model parameter $\Lambda$.}\label{fig:width2}
\end{figure}

Different from $T_{c\bar{s}1}(2460)$, all the considered decay processes of $T_{c\bar{s}1}^{a+}(2470)$, including $T_{c\bar{s}1}^{a+}(2470) \to D_{s}^{\ast+} \pi^0$, $T_{c\bar{s}1}^{a+}(2470) \to D_{s0}^{\ast+} \pi^0$ and $T_{c\bar{s}1}^{a+}(2470) \to T_{c\bar{s}0}^{a} \pi$, are isospin-conserving allowed processes, leading to much larger partial decay widths. 

In Fig.~\ref{fig:width2}, we present our estimations of the partial widths of processes $T_{c\bar{s}1}^{a+}\to D_{s}^{*+}\pi^{0}$, $D_{s0}^{*+}\pi^{0}$, and $T_{c\bar{s}0}^{a}\pi$ as functions of the model parameter $\Lambda$. For the $T_{c\bar{s}1}^{a+}\to D_{s}^{*+}\pi^{0}$ process, our estimations indicate that the  partial width increases rather rapidly with the increasing of model parameter $\Lambda$. In particular, the partial width increases from $13.35$ to $196.1~\mathrm{MeV}$ when $\Lambda$ increases from $1~\mathrm{GeV}$ to $3~\mathrm{GeV}$, which is about $10^3$ times the partial width of $T_{c\bar{s}1}^{f+}\to D_{s}^{*+}\pi^{0}$. As for the decay processes $T_{c\bar{s}1}^{a+}(2470) \to D_{s0}^{\ast+} \pi^0$ and $T_{c\bar{s}1}^{a+}(2470) \to T_{c\bar{s}0}^{a} \pi$, they are suppressed by kinematic limitation. The partial decay widths for $T_{c\bar{s}1}^{a+}\to D_{s0}^{*+}\pi^{0}$ and $T_{c\bar{s}1}^{a+}\to T_{c\bar{s}0}^{a}\pi$ increase from $0.18~\mathrm{MeV}$ to $0.20~\mathrm{MeV}$ and from $0.10~\mathrm{MeV}$ to $0.11~\mathrm{MeV}$, respectively.

\subsection{Strong decays of $T_{\bar{c}\bar{s}1}^{f0}$ and $T_{\bar{c}\bar{s}1}^{a0}$}


Regarding $T_{\bar{c}\bar{s}1}^{a0}$ and $T_{\bar{c}\bar{s}1}^{f0}$, their decays into charm-strange and light mesons or charm and strange mesons are forbidden due to their quark components, $J^P$ quantum numbers conservation and kinematic limitations. The possible strong decay processes are $T_{\bar{c}\bar{s}1}^{a0}\to T_{\bar{c}\bar{s}0}^{f0}\pi^{0}$, $T_{\bar{c}\bar{s}1}^{a0}\to T_{\bar{c}\bar{s}0}^{a0}\pi^{0}$, $T_{\bar{c}\bar{s}1}^{f0}\to T_{\bar{c}\bar{s}0}^{f0}\pi^{0}$, and $T_{\bar{c}\bar{s}1}^{f0}\to T_{\bar{c}\bar{s}0}^{a0}\pi^{0}$ if there exist $\bar{D}K$ molecular states with $I=0$ and $I=1$. Similar to the transitions between $D^{*}K$ and $DK$ molecules, the partial decay widths for $T_{\bar{c}\bar{s}1}^{a0}\to T_{\bar{c}\bar{s}0}^{f0}\pi^{0}$, $T_{\bar{c}\bar{s}0}^{a0}\pi^{0}$, and $T_{\bar{c}\bar{s}1}^{f0}\to T_{\bar{c}\bar{s}0}^{a0}\pi^{0}$ are approximately several hundred $\mathrm{keV}$ since these processes are only suppressed by kinematic limitations. While for the isospin-violating process $T_{\bar{c}\bar{s}1}^{f0}\to T_{\bar{c}\bar{s}0}^{f0}\pi^{0}$, the partial decay width is expected to be only about $0.1~\mathrm{keV}$ due to the suppressions of both isospin violation and kinematic limitation.

\section{Summary}\label{sec4}
In the past two decades, a series of tetraquark candidates, including $D_{s0}^{\ast}(2317)$, $T_{c\bar{s}0}(2327)$, $D_{s1}(2460)$, $X_{0,1}(2900)$ (now named $T_{\bar{c}\bar{s}0}^{\ast} (2870)/T_{\bar{c}\bar{s}1}^{\ast} (2900)$), and $T_{c\bar{s}0}^{a++,0}(2900)$, have been experimentally reported. All these states are very close the thresholds of $D^{(\ast)} K^{(\ast)}$, leading to the prosperity of molecular interpretations for these states. Near the threshold of $D^\ast K/\bar{D}^\ast K$, one molecular candidate has been observed, which is $D_{s1}(2460)$ (in the present estimations, it is called $T_{c\bar{s}1}^{f}(2460)$ following the LHCb naming convention), Similar to the case of $D_{s0}^{\ast}(2317)$ and $T_{c\bar{s}0}(2327)$ near the $DK$ threshold, an isospin partner of $T_{c\bar{s}1}^{f}(2460)$, named $T_{c\bar{s}1}^{a}$ is expected to exist near the $D^\ast K$ threshold. Moreover, the observations of  $T_{\bar{c}\bar{s}0}^{\ast} (2870)$ and $T_{c\bar{s}0}^{a++,0}(2900)$ may also indicate the existence of molecular states composed of $\bar{D}^\ast K$ named $T_{\bar{c}\bar{s}1}^{f}$ and $T_{\bar{c}\bar{s}1}^{a}$.

In the present work, we investigate the decay properties of these possible $D^\ast K$ and $\bar{D}^\ast K $ molecular states by employing an effective Lagrangian approach. The decay processes $T_{c\bar{s}1}^{f/a+}\to D_{s}^{*+}\pi^{0}$, $D_{s0}^{*+}\pi^{0}$, $T_{c\bar{s}0}^{a+}\pi^{0}$, $T_{\bar{c}\bar{s}1}^{f/a0}\to T_{\bar{c}\bar{s}0}^{f0}\pi^{0}$, $T_{\bar{c}\bar{s}0}^{a0}\pi^{0}$ have been considered. Our estimates indicate that the decay processes of $T_{c\bar{s}1}^{f+}(2460)$ that we have considered in the present article,  are suppressed by isospin violation or kinematic limitation or both, resulting in a very small partial width. As for $T_{c\bar{s}1}^{a+}(2470)$,  the decay process $T_{c\bar{s}1}^{a+}(2470) \to D_s^{\ast+} \pi^0$ is isospin conserved, in the considered parameter range, its partial width varies from 13.35 MeV to near 200 MeV, while the pionic transitions between $T_{c\bar{s}1}^{a+}(2470)$ and $D_{s0}^{\ast+}(2317)/T_{c\bar{s}0}(2327)$ are suppressed by kinematic limitation, and the partial widths are estimated to be of order 0.1 MeV. As for $T_{\bar{c}\bar{s}1}^{f/a}$, they can only decay into $T_{\bar{c}\bar{s} 0}^{f/a} \pi^0$ if there exist $\bar{D}K$ molecular states. The present estimations indicate that the widths of these states should be of order $10^2$ keV.

\section{Acknowledgments}
This study is partly supported by the National Natural Science Foundation of China under Grant Nos. 12175037 and 12335001, and is supported, in part, by the National Key Research and Development Program under Contract No. 2024YFA1610503. Zi-Li Yue is also supported by the SEU Innovation Capability Enhancement Plan for Doctoral Students (Grant No. CXJH$\_$SEU 24135) and the China Scholarship Council (Grant No. 202406090305).


\bibliographystyle{apsrev4-1}
\bibliography{DstarK.bib}

\end{document}